\documentclass[times,twocolumn,final,longtitle]{elsarticle}

\usepackage{medima}
\usepackage{framed,multirow}
\usepackage{multirow}%
\usepackage{subcaption} 
\usepackage{booktabs}%
\usepackage{lipsum}
\usepackage{amsmath} 

\usepackage[export]{adjustbox}
\usepackage{float}
\usepackage{dblfloatfix}
\usepackage{makecell}
\usepackage{graphicx}
\usepackage{amssymb}
\usepackage{amsfonts}

\usepackage{amssymb}
\usepackage{latexsym}

\usepackage{url}

\usepackage{hyperref}

\definecolor{newcolor}{rgb}{.8,.349,.1}

\journal{Medical Image Analysis}

\begin{document}

\verso{The AortaSeg24 Challenge}

\begin{frontmatter}

\title{Multi-Class Segmentation of Aortic Branches and Zones in Computed Tomography Angiography: The AortaSeg24 Challenge}

\author[1]{Muhammad Imran}
\author[2]{Jonathan R. Krebs}
\author[3]{Vishal Balaji Sivaraman}
\author[3]{Teng Zhang}
\author[4]{Amarjeet Kumar}
\author[2]{Walker R. Ueland}
\author[2]{Michael J. Fassler}
\author[5]{Jinlong Huang}
\author[5]{Xiao Sun}
\author[5]{Lisheng Wang}
\author[6]{Pengcheng Shi}
\author[7]{Maximilian Rokuss}
\author[7]{Michael Baumgartner}
\author[7]{Yannick Kirchhof}
\author[7]{Klaus H. Maier-Hein}
\author[7]{Fabian Isensee}
\author[8]{Shuolin Liu}
\author[8]{Bing Han}
\author[9]{Bong Thanh Nguyen}
\author[{9,101}]{Dong-jin Shin}
\author[{9,101}]{Park Ji-Woo}
\author[{9,101}]{Mathew Choi}
\author[{9,101}]{Kwang-Hyun Uhm}
\author[{9,101}]{Sung-Jea Ko}
\author[10]{Chanwoong Lee}
\author[10]{Jaehee Chun}
\author[10]{Jin Sung Kim}
\author[11]{Minghui Zhang}
\author[11]{Hanxiao Zhang}
\author[11]{Xin You}
\author[11]{Yun Gu}
\author[12]{Zhaohong Pan}
\author[12]{Xuan Liu}
\author[12]{Xiaokun Liang}
\author[13]{Markus Tiefenthaler}
\author[13]{Enrique Almar-Munoz}
\author[13]{Matthias Schwab}
\author[14]{Mikhail Kotyushev}
\author[14]{Rostislav Epifanov}
\author[{15a,15b}]{Marek Wodzinski}
\author[{15b,15c,15d}]{Henning Müller}
\author[16]{Abdul Qayyum}
\author[16]{Moona Mazher}
\author[16]{Steven A. Niederer}
\author[17]{Zhiwei Wang}
\author[17]{Kaixiang Yang}
\author[18]{Jintao Ren}
\author[18]{Stine Sofia Korreman}
\author[19]{Yuchong Gao}
\author[19]{Hongye Zeng}
\author[19]{Haoyu Zheng}
\author[19]{Rui Zheng}
\author[20]{Jinghua Yue}
\author[20]{Fugen Zhou}
\author[20]{Bo Liu}
\author[4]{Alexander Cosman}
\author[22]{Muxuan Liang}
\author[24]{Chang Zhao}
\author[2]{Gilbert R. Upchurch, Jr.}
\author[23]{Jun Ma}
\author[21]{Yuyin Zhou}
\author[2]{Michol A. Cooper\fnref{fn1}}
\author[1,3]{Wei Shao\corref{cor1}\fnref{fn1}}
\cortext[cor1]{Corresponding author. \\
    \makebox[0pt][l]{\hspace{2em} E-mail: weishao@ufl.edu (W. Shao)} \\
    \makebox[0pt][l]{\hspace{2em} ORCID: 0000-0003-4931-4839 (W. Shao)}}
\fntext[fn1]{Equal contribution as senior authors.}

\address[1]{Department of Medicine, University of Florida, Gainesville, FL, 32611, United States}
\address[2]{Department of Surgery, University of Florida, Gainesville, FL, 32611, United States}
\address[3]{Department of Electrical and Computer Engineering, University of Florida, Gainesville, FL, 32611, United States}
\address[4]{Department of Computer, Information Science, and Engineering, University of Florida, Gainesville, FL, 32611, United States}
\address[5]{Shanghai Jiao Tong University, Shanhai, China}
\address[6]{Electronic \& Information Engineering School, Harbin Institute of Technology (Shenzhen), Shenzhen, China}
\address[7]{Division of Medical Image Computing, German Cancer Research Center (DKFZ), Heidelberg, Germany}
\address[8]{CANON MEDICAL SYSTEMS (CHINA) CO., LTD, Beijing, China}
\address[9]{MedAI, Seoul, South Korea}
\address[101]{Korea University, Seoul, South Korea}
\address[10]{Department of Radiation Oncology, Yonsei Cancer Center, Heavy Ion Therapy Research Institute, Yonsei University College of Medicine, Seoul, Korea}
\address[11]{Institute of Medical Robotics, Shanghai Jiao Tong University, Shanghai, China}
\address[12]{Shenzhen Institute of Advanced Technology, Chinese Academy of Sciences, Shenzhen, 518055, China }
\address[13]{Medical University of Innsbruck, Innsbruck, Austria}
\address[14]{Novosibirsk State University, Novosibirsk, Russia}
\address[15a]{AGH University of Krakow, Krakow, Poland}
\address[15b]{Institute of Informatics, University of Applied Sciences Western Switzerland (HES-SO), Sierre, Switzerland}
\address[15c]{University of Geneva, Geneva, Switzerland}
\address[15d]{The Sense Innovation and Research Center, Sion, Switzerland}
\address[16]{National Heart and Lung Institute, Faculty of Medicine, Imperial College London, London, United Kingdom}
\address[17]{ Wuhan National Laboratory for Optoelectronics, Huazhong University of Science and Technology, Wuhan, Hubei, China}
\address[18]{Aarhus University, Department of Clinical Medicine, Nordre Palle Juul-Jensens Blvd. 11, 8200 Aarhus, Denmark}
\address[19]{Shanghaitech University, Pudong, Shanghai, China}
\address[20]{Image Processing Center, Beihang University, Beijing, China}
\address[21]{Department of Computer Science and Engineering, University of California, Santa Cruz, CA, 95064, United States}
\address[22]{Department of Biostatistics, University of Florida, Gainesville, FL, 32611, United States}
\address[23]{University Health Network, Ontario, Canada}
\address[24]{Agronomy Department, University of Florida, Gainesville, FL, 32611, United States}


\begin{abstract}
Multi-class segmentation of the aorta in computed tomography angiography (CTA) scans is essential for diagnosing and planning complex endovascular treatments for patients with aortic dissections. However, existing methods reduce aortic segmentation to a binary problem, limiting their ability to measure diameters across different branches and zones. Furthermore, no open-source dataset is currently available to support the development of multi-class aortic segmentation methods. To address this gap, we organized the AortaSeg24 MICCAI Challenge, introducing the first dataset of 100 CTA volumes annotated for 23 clinically relevant aortic branches and zones. This dataset was designed to facilitate both model development and validation. The challenge attracted 121 teams worldwide, with participants leveraging state-of-the-art frameworks such as nnU-Net and exploring novel techniques, including cascaded models, data augmentation strategies, and custom loss functions. We evaluated the submitted algorithms using the Dice Similarity Coefficient (DSC) and Normalized Surface Distance (NSD), highlighting the approaches adopted by the top five performing teams. This paper presents the challenge design, dataset details, evaluation metrics, and an in-depth analysis of the top-performing algorithms. The annotated dataset, evaluation code, and implementations of the leading methods are publicly available to support further research. All resources can be accessed at \url{https://aortaseg24.grand-challenge.org.}
\end{abstract}


\end{frontmatter}

\section{Introduction}
\label{sect:introduction}
The aorta, the body’s largest artery, carries oxygen-rich blood from the heart to the head, neck, limbs, and lower body. Aortic dissections, caused by a tear in the aortic wall, are life-threatening conditions that disrupt blood flow, compromise organ perfusion, and may lead to complications such as rupture~\citep{juraszek2022update, yin2023research, carrel2023acute, rolf2024mechanisms}. Advances in medical imaging, particularly computed tomography angiography (CTA), and therapies like endovascular stent grafts have revolutionized the diagnosis and treatment of aortic dissections~\citep{rolf2024mechanisms}. A key step in leveraging these advancements is the segmentation of the aorta into clinically relevant branches and  Society for Vascular Surgery/Society of Thoracic Surgeons (SVS/STS) zones~\citep{lombardi2020society} (see Fig.~\ref{fig:aorta-branches}). This segmentation method enables automated measurement of the diameters and volumes of different aortic branches and zones, which are essential for accurate device selection, optimal stent placement, and improved outcomes while minimizing complications~\citep{carrel2023acute, rolf2024mechanisms}.

Despite the critical need for multi-class aortic segmentation, existing methods often reduce the problem to binary segmentation, treating the aorta as a single object~\citep{cao2019fully, lyu2021dissected, chen2022deep}. This simplification overlooks the anatomical complexity and clinical importance of differentiating aortic zones and branches. Consequently, these methods have limited utility for tasks such as measuring zonal diameters and volumes, assessing the extent of dissections, and guiding precise device placement. Furthermore, no open-source dataset currently exists for multi-class segmentation of the aorta, its branches, and zones. Most available datasets are small, proprietary, and focused solely on binary segmentation, restricting the development and validation of advanced segmentation algorithms capable of addressing real-world clinical challenges. 

To overcome these limitations, it is essential to identify and address the unique challenges associated with multi-class segmentation of the aorta:
\begin{figure}[!hbt]%
\centering
\includegraphics[width=0.3\textwidth]{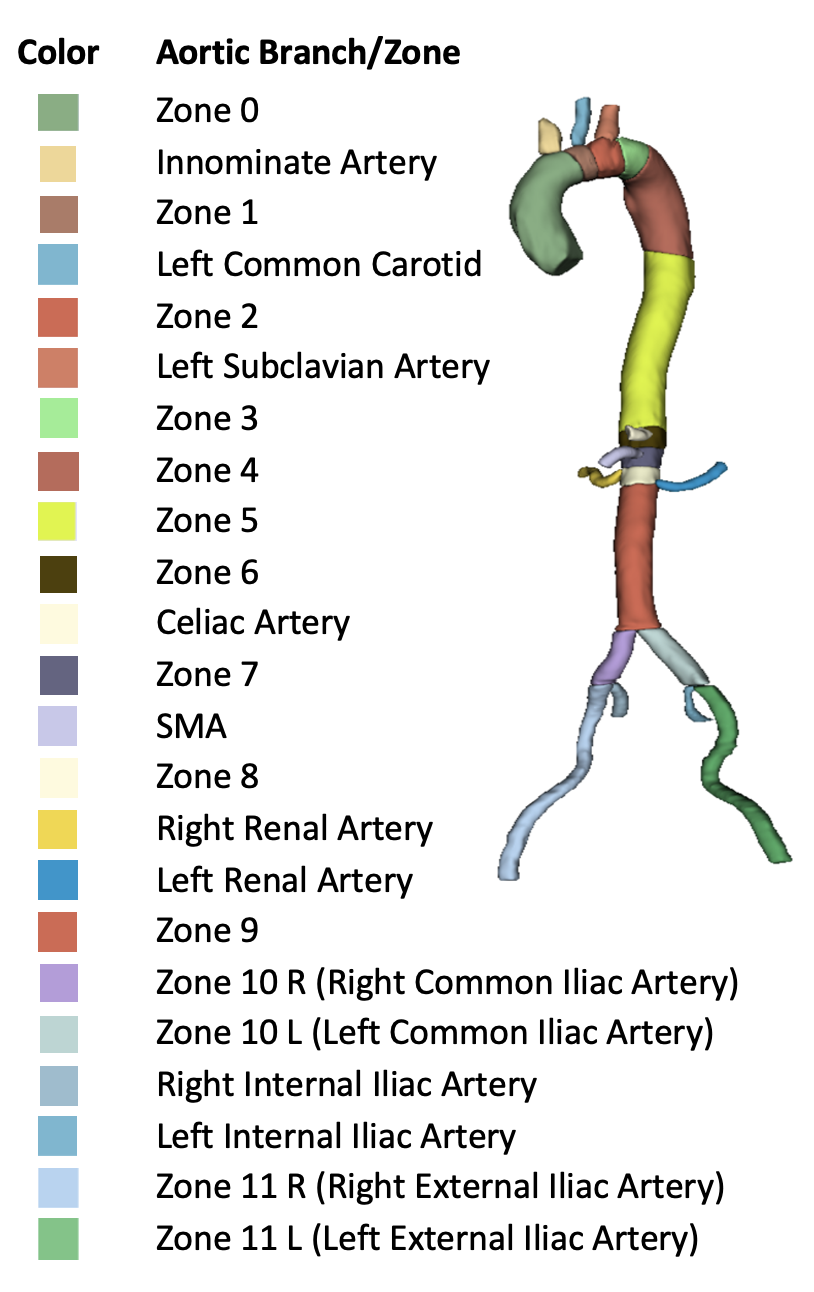}
\caption{Schematic representation of different aortic branches and zones.}
\label{fig:aorta-branches}%
\end{figure}
\begin{enumerate}
    \item \textbf{Complex Anatomy:}  The aorta has multiple zones and branches, each with distinct shapes, orientations, and sizes. These features vary significantly across patients due to differences in age, sex, body size, and disease states. 
    \item \textbf{High Annotation Effort:} A single 3D CTA volume often contains 500-800 2D slices, making manual annotation of the aorta and its branches time-consuming. Moreover, expert clinician input is required to ensure annotation accuracy and consistency, further increasing the labor involved in creating high-quality datasets.
    \item \textbf{Limited Dataset Availability:} Existing datasets are typically small, proprietary, and focused on binary segmentation, which constrains the development of multi-class aortic segmentation algorithms.
    \item \textbf{Lack of Standardized Benchmarking:} The absence of standardized benchmarks for evaluating multi-class segmentation methods limits the ability to objectively compare algorithms and adopt them in clinical practice.
\end{enumerate}

These challenges highlight the critical need for initiatives that foster innovation and enhance the accuracy and clinical utility of aortic segmentation methods. To bridge these gaps, we launched the AortaSeg24 challenge, which was accepted by the International Conference on Medical Image Computing and Computer-Assisted Intervention (MICCAI) in 2024. As a cornerstone of this initiative, we developed the first large-scale dataset comprising 100 CTA volumes, each annotated by experts to delineate 23 aortic branches and SVS/STS zones. This dataset not only represents a significant advancement in the availability of resources for aortic segmentation but also provides a unique opportunity for researchers to design and evaluate cutting-edge algorithms for accurate, automated, multi-class segmentation. We summarize our key contributions as follows:

\begin{enumerate}
    \item \textbf{Developing the AortaSeg24 Dataset:} We created a dataset of 100 CTA volume images with expert annotations of 23 aortic branches and SVS/STS zones, addressing a critical gap in the availability of publicly accessible datasets. 
    \item \textbf{Hosting the AortaSeg24 Challenge:} We organized the AortaSeg24 MICCAI challenge to provide a standardized platform for developing and evaluating innovative approaches to multi-class aortic segmentation. 
    \item \textbf{Open Access and Clinical Utility:} To promote reproducibility and future research, we made the dataset, evaluation code, and top-performing methods publicly available. These resources support clinical applications by enabling precise aortic measurements, optimal device selection, and enhanced treatment planning. 
\end{enumerate}

In this paper, we present the outcomes of the AortaSeg24 Challenge, including a detailed description of the dataset, evaluation criteria, and the top-performing algorithms. We also analyze the strengths and limitations of these methods. By setting a new standard in aortic segmentation, this work aims to bridge the gap between research innovation and clinical utility, paving the way for more effective treatments for aortic diseases.

\section{Related Work}
\label{sect:prior-work}
\subsection{Existing Aortic Segmentation Methods}
\label{sect:related-work}

The aorta can be divided into the ascending aorta, aortic arch, thoracic aorta, abdominal aorta, and iliac arteries. Most existing methods focus on binary segmentation of either the thoracic aorta or the entire aorta, without distinguishing between different aortic branches and zones~\citep{lyu2021dissected, lin2023deformable, de2022quantification, sieren2022automated, comelli2021deep, zhao2022segmentation}. While most studies utilize 3D CT or CTA images, some have explored aortic segmentation in other modalities, such as 3D MRI~\cite{manokaran2023fully, guo2024deep, berhane2020fully, marin20234d} and 3D ultrasound~\citep{maas2024automatic}. Additionally, certain methods address the segmentation of the true and false lumen in aortic dissections. For instance, in~\citep{jung2024zozi} the ZOZI-seg model was introduced, a two-stage framework that combines a 3D transformer with a 3D U-Net for localized texture refinement. This model, trained on private data, segments the aorta into the true lumen, false lumen, and thrombosis. Similarly, ~\cite{mu2023automatic} proposed a method that segments the aorta into the true lumen and intraluminal thrombosis, leveraging a private dataset of 70 3D CTA volumes. 

Recently, a few techniques have gone beyond binary segmentation to multi-class segmentation, although they remain focused on chest CTA. For example, in~\citep{zhong2021segmentation}, a U-Net model with attention gating was developed for multi-class segmentation of the thoracic aorta using a private CTA dataset. Similarly, researchers in~\citep{koo2024deep} proposed a CNN-based segmentation model trained on a private dataset of 30 3D chest CTA volumes to segment the aorta (excluding the iliac arteries) into nine zones using a multi-atlas method. As summarized in Table~\ref{tab:literature_review}, existing methods are limited to either binary segmentation of the aorta or multi-class segmentation focused on the thoracic aorta. This highlights a critical unmet need for a large, standardized, open-source dataset enabling multi-class segmentation of the entire aorta, its branches, and zones, from the ascending aorta to the iliac arteries.

\begin{table*}[!hbt]
\caption{Summary of existing aortic segmentation studies. CT: computed tomography; CTA: computed tomography angiography; MRI: magnetic resonance imaging; TL: true lumen; FL: false lumen.}
\centering
\renewcommand{\arraystretch}{1.2} 
\footnotesize
\begin{tabular}{|c|c|c|c|c|c|}
\hline
\textbf{Model} & \textbf{Modality} & \textbf{Image Dimension} & \textbf{Dataset Type} & \textbf{Segmented Object(s)} & \textbf{\# of Image Volumes} \\ \hline
\cite{cao2019fully}          & CTA            & 3D          & Private        & TL and FL                     & 276 \\ \hline
\cite{jung2024zozi}          & CT             & 3D          & Private        & TL, FL, and thrombosis         & 253 \\ \hline
\cite{mu2023automatic}       & CTA            & 3D          & Private        & TL and thrombosis              & 70  \\ \hline
\cite{lyu2021dissected}      & CTA            & 3D          & Private        & Aorta                          & 42  \\ \hline
\cite{koo2024deep}           & CT             & 3D          & Private        & Nine thoracic aortic zones     & 704 \\ \hline
\cite{chen2021multi}         & CTA            & 3D          & Private        & TL and FL                      & 120 \\ \hline
\cite{xu2023sliceprop}       & CTA            & 3D          & Public         & TL and FL                      & 33  \\ \hline
\cite{li2023evaluating}      & CTA            & 3D          & Public         & TL and FL                      & 100 \\ \hline
\cite{10.1007/978-3-031-53241-2_7} & CTA      & 3D          & Public         & Aorta                          & 38  \\ \hline
\cite{cheng2020deep}         & CT             & 2D          & Private        & TL and FL                      & 20  \\ \hline
\cite{noothout2018automatic} & CT             & 3D          & Private        & Three thoracic aortic zones    & 24  \\ \hline
\cite{zhong2021segmentation} & CTA            & 3D          & Private        & Six thoracic aortic zones      & 194 \\ \hline
\cite{manokaran2023fully}    & 4D flow MRI    & 3D          & Private        & Aorta                          & 114 \\ \hline
\cite{feiger2021evaluation}  & CTA            & 2D and 3D   & Private        & TL and FL                      & 21  \\ \hline
\cite{lin2023deformable}     & CT             & 3D          & Private        & TL and FL                      & 267 \\ \hline
\cite{sun2024paratranscnn}   & CTA            & 3D          & Public         & Aorta                          & 56  \\ \hline
\cite{hahn2020ct}            & CTA            & 3D          & Private        & TL and FL                      & 153 \\ \hline
\cite{yu2021three}           & CTA            & 3D          & Private        & TL and FL                      & 139 \\ \hline
\cite{guo2024deep}           & MRI            & 3D          & Private        & Thoracic aorta                 & 391 \\ \hline
\cite{feng2023automatic}     & CTA            & 3D          & Private        & TL and FL                      & 463 \\ \hline
\cite{sieren2022automated}   & CTA            & 3D          & Private        & Aorta                          & 191 \\ \hline
\cite{comelli2021deep}       & CTA            & 3D          & Private        & Thoracic aorta                 & 72  \\ \hline
\end{tabular}
\label{tab:literature_review}
\end{table*}

\subsection{Existing Aortic Segmentation Datasets}
\label{sect:about-the-dataset}
\begin{figure*}[!hbt]%
\centering
\includegraphics[width=0.75\textwidth]{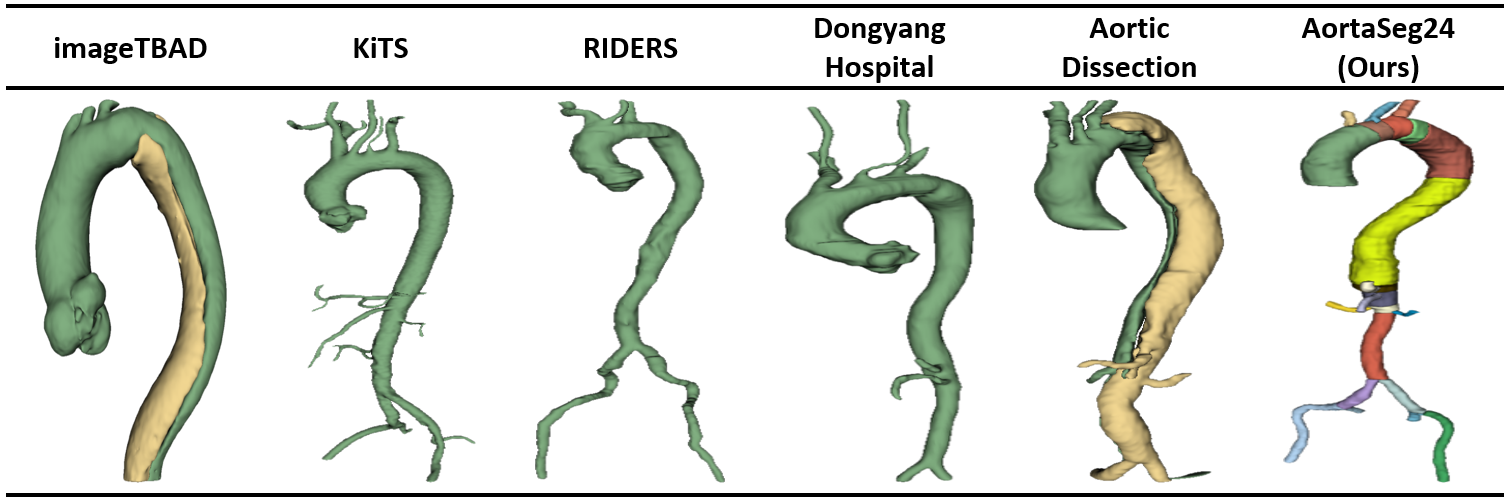}
\caption{Comparison of various publicly available aortic segmentation datasets}
\label{fig:aorta-datasets-for-comparison}%
\end{figure*}

As summarized in Table~\ref{tab:literature_review}, most studies rely on proprietary datasets that are not publicly accessible. However, a few publicly available datasets exist for aortic segmentation. The ImageTBAD dataset~\citep{yao2021imagetbad} contains annotations for 100 CTA images from patients with type B aortic dissections treated at the Guangdong Provincial People’s Hospital in China. Of these, 32 images are annotated for true lumen and false lumen, while the remaining 68 images include annotations for true lumen, false lumen, and false lumen thrombus.  Despite its value, this dataset has notable limitations: it covers only a portion of the aorta (Figure~\ref{fig:aorta-datasets-for-comparison}(a)) and focuses solely on aortic dissections, omitting branches and zones critical for monitoring zonal progression. The AVT dataset~\citep{radl2022avt} is a publicly available multi-center dataset of 56 CTA images from three sources: 20 cases from the KiTS19 Grand Challenge~\cite{heller2019kits19, heller2021state}, 18 from the Rider Lund CT dataset~\citep{zhao2015data}, and 18 from Dongyang Hospital. Images from KiTS and Dongyang hospital are free of pathologies, while those from the Rider dataset include aortic dissections and abdominal aortic aneurysms. Unlike ImageTBAD, the AVT dataset spans the entire aorta and its branches. However, it annotates the aorta and all its branches as a single label, without distinguishing individual branches and zones (Figure~\ref{fig:aorta-datasets-for-comparison} (a)-(c)). The recently released Aortic Dissection Dataset~\citep{mayer2024type} comprises 40 CTA images. Unlike the ImageTBAD dataset, it includes the entire aorta and some aortic branches in most cases (Figure \ref{fig:aorta-datasets-for-comparison}) but provides labels only for true lumen and false lumen, omitting individual branches and zones.

\subsection{Relevant Image Segmentation Challenges}
\label{sect:aorta-related-challenges}
Medical image segmentation has been the focus of numerous medical imaging challenges, including notable ones such as the Head and Neck Organ-at-Risk CT \& MR Segmentation Challenge (HaN-Seg)~\citep{podobnik2024han}, the Fast and Low-Resource Semi-Supervised Abdominal Organ Segmentation in CT (FLARE 2022)~\citep{flare2022fast}, the Segmentation of Organs-at-Risk and Gross Tumor Volume of NPC for Radiotherapy Planning (SegRap2023)~\citep{luo2023segrap2023}, and the KiTS21 Challenge~\citep{heller2023kits21}. Despite their contributions, none addressed the segmentation of the aorta. The 2023 Segmentation of the Aorta (SEG.A.2023) challenge~\citep{pepe2024segmentation} aimed to advance methods for automated binary segmentation of the aortic vessel tree in CTA images. The challenge provided participants with a dataset of 56 annotated CTA images from three institutions to develop their models. Although SEG.A.2023 marked significant progress, it was limited to binary segmentation of the aorta versus non-aorta and did not include detailed annotations for aortic branches or zones.

\section{Methods}
\subsection{Challenge Organization}
\label{sect:challenge-setup-details}

The AortaSeg24 grand challenge, hosted by the University of Florida and accepted by the 2024 Medical Image Computing and Computer-Assisted Intervention (MICCAI) conference, adhered to rigorous ethical and scientific standards. We secured Institutional Review Board (IRB) approval, conducted thorough risk assessments, and implemented strict data usage agreements (DUAs) to protect patients' privacy. In addition to possessing verified Grand Challenge accounts, participants were required to provide real names and affiliations for their DUAs. This enabled challenge administrators to vet and validate the identities of prospective participants.

The challenge, which opened on May 9, 2024, consisted of three phases: development, validation, and testing. Approved participants from around the world, each limited to joining only one team, were provided with 50 annotated 3D CTA images via the University of Florida Dropbox to develop and train their segmentation models. During the development phase, teams were prohibited from using external training data and were given one month to familiarize themselves with the dataset and train their models. Starting on June 6, 2024, participants could upload their Docker images to the challenge website for validation on 10 hidden CTA images. Detailed instructions, including a baseline 3D U-Net model and code for computing Dice scores, were shared on GitHub (\url{https://github.com/mirthAI/AortaSeg24}). Participants received immediate feedback on Dice scores and were allowed up to ten attempts to refine their models during the validation phase, which concluded on August 23, 2024. Top-performing teams were invited to the testing phase, beginning on September 18, 2024, where they submitted Docker containers to evaluate their models on 40 hidden test cases. These test cases remained entirely unseen by participants throughout the challenge. The testing phase concluded on September 30, 2024, marking the end of the competition.

 Top-performing teams received multiple awards: the top five teams received cash prizes, the top ten received honorable mention certificates, and those fulfilling co-authorship criteria were invited to contribute to a challenge review paper. The co-authorship criteria included making code publicly available on GitHub, providing instructions for reproducibility, and surpassing the baseline 3D U-Net model. In this way, the AortaSeg24 challenge fostered a collaborative research environment, encouraged the development and dissemination of improved aortic segmentation models, and upheld the highest standards of academic integrity, patient data protection, and methodological rigor.

\subsection{Challenge Dataset}
\label{sect:aortaseg24-dataset}
\subsubsection{Image Acquisition}
\label{sect:image-acquisition}
We conducted a retrospective review of patients diagnosed with uncomplicated type B aortic dissection (uTBAD) between October 2011 and March 2020, using a prospectively maintained institutional database (IRB approval and consent waiver: 202100962). We defined uTBAD as the absence of malperfusion, rupture, rapid degeneration, or refractory pain. All uTBAD patients were medically managed without surgical intervention during their initial admission. Patients with type A dissection, intramural hematoma, penetrating aortic ulcer, chronic aortic dissection, or those who underwent TEVAR during the index admission were excluded.

Standard CTA imaging consists of three phases: a non-contrast phase, an arterial phase, and a delayed phase. The non-contrast phase is used to detect hematomas or plaques that may be obscured by contrast. This is followed by the arterial phase, where iodinated contrast medium (ICM) is rapidly injected to optimize arterial visualization and is synchronized with peak aortic contrast arrival. The delayed phase is carefully timed to evaluate slow-filling and venous structures. Given the variability in protocols from local imaging centers, our study only included cases in which CTA images had a slice thickness of $\leq 3$ mm at both diagnosis and beyond three months from discharge.

In total, our dataset includes 100 CTA image volumes from 100 uTBAD patients. The original image sizes range from $512 \times 512 \times 245$ to $512 \times 512 \times 962$, with an average of 696 slices per volume. The in-plane resolution varies from $0.689 \times 0.689$ mm to $1 \times 1$ mm, averaging $0.863 \times 0.863$ mm. The through-plane resolution (slice thickness) ranges from $0.8$ mm to $2.52$ mm, with an average of $1.03$ mm. To standardize the dataset and eliminate heterogeneity, we resampled the image volumes to an isotropic spacing of $1 \times 1 \times 1$ mm. After resampling, the minimum, maximum, and average image sizes were $353 \times 353 \times 500$, $516 \times 516 \times 801$, and $441 \times 441 \times 686$, respectively.

\subsubsection{Clinically Relevant Aortic Zones and Branches}
\label{sect:segmentation-overview}
The classification of aortic zones is crucial for planning endovascular interventions. According to the Society for Vascular Surgery (SVS) and the Society of Thoracic Surgeons (STS)~\citep{lombardi2020society}, the aorta is divided into specific zones based on its branches. This study focuses on segmenting the clinically relevant branches outlined in Table~\ref{tab:aortic_branches_zones}, which serve as the basis for defining SVS/STS zones (see Table~\ref{tab:aortic_branches_zones}). These zones segment the aorta into regions critical for diagnosis and treatment planning.

\begin{table*}[ht]
\small 
\centering
\caption{Description of Clinically Relevant Aortic Branches and Zones}
\label{tab:aortic_branches_zones}
\begin{adjustbox}{width=\textwidth}
\begin{tabular}{|l|p{14cm}|}
\hline
\textbf{Branch/Zone}         & \textbf{Description} \\ \hline
Innominate artery               & First branch of the aortic arch, supplying blood to the right arm and the right side of the head and neck. \\ \hline
Left common carotid artery      & Second branch of the aortic arch, supplying blood to the left side of the head and neck. \\ \hline
Left subclavian artery          & Third branch of the aortic arch, supplying blood to the left upper limb, neck, and chest wall. \\ \hline
Celiac artery                   & Supplies blood to the stomach, liver, spleen, and pancreas. \\ \hline
Superior mesenteric artery      & Supplies blood to the small intestine, cecum, ascending colon, and part of the transverse colon. \\ \hline
Left renal artery               & A branch of the abdominal aorta that supplies oxygenated blood to the left kidney. \\ \hline
Right renal artery              & A branch of the abdominal aorta that supplies oxygenated blood to the right kidney. \\ \hline
Left common iliac artery/ Zone 10 L        & A terminal branch of the abdominal aorta that supplies blood to the left lower limb and pelvic region, dividing into the internal and external iliac arteries. \\ \hline
Right common iliac artery/Zone 10 R       & A terminal branch of the abdominal aorta that supplies blood to the right lower limb and pelvic region, dividing into the internal and external iliac arteries. \\ \hline
Left external iliac artery/Zone 11 L      & A branch of the left common iliac artery that supplies blood to the left lower limb. \\ \hline
Right external iliac artery/Zone 11 R     & A branch of the right common iliac artery that supplies blood to the right lower limb. \\ \hline
Left internal iliac artery      & A branch of the left common iliac artery that supplies blood to the pelvic organs, gluteal region, and perineum. \\ \hline
Right internal iliac artery     & A branch of the right common iliac artery that supplies blood to the pelvic organs, gluteal region, and perineum. \\ \hline
Zone 0                          & From the aortic root to the innominate artery. \\ \hline
Zone 1                          & From the innominate artery to the left common carotid artery. \\ \hline
Zone 2                          & From the left common carotid artery to the left subclavian artery. \\ \hline
Zone 3                          & From the left subclavian artery to 2 cm distal to it. \\ \hline
Zone 4                          & From the distal end of Zone 3 to the T6 vertebral body. \\ \hline
Zone 5                          & From the distal end of Zone 4 to the celiac artery. \\ \hline
Zone 6                          & From the celiac artery to the superior mesenteric artery. \\ \hline
Zone 7                          & From the superior mesenteric artery to the most proximal renal artery. \\ \hline
Zone 8                          & From the most proximal renal artery to the most distal renal artery. \\ \hline
Zone 9                          & From the most distal renal artery to the bifurcation of the aorta into the left and right common iliac arteries. \\ \hline
\end{tabular}
\end{adjustbox}
\end{table*}

\subsubsection{Data Annotation}
\label{sect:data-annotation}
Our expert vascular surgeon (MC), who has more than 10 years of experience interpreting CTA images of the chest, abdomen, and pelvis, trained a team of four (MI, VSB, TZ, and AK) to identify critical aortic structures on CTA images. After the training, the trainees evenly split the 100 CTA images and performed manual annotation of the aorta, its branches, and zones on 3D CT images using the 3D Slicer tool~\citep{fedorov20123d}. The annotation process was divided into two phases: first annotating the aorta and its thirteen branches, then dividing the aorta into zones based on the branch annotations. Throughout the annotation process, we leveraged the complementary information from the axial, sagittal, and coronal views of the CT scans to ensure precision.

For the segmentation of the aorta and the six iliac artery branches, we primarily relied on the axial view, referring to the sagittal and coronal views as needed. The celiac artery and superior mesenteric artery were primarily segmented on sagittal slices, while the left and right renal arteries were primarily segmented on coronal slices. To ensure consistency, each branch segmentation ended at the bifurcation point where it split into multiple branches. For example, the segmentation of the innominate artery ended at the point where it branches into the right subclavian artery and the right common carotid artery.

Following the segmentation of the aorta and its branches, the aorta was divided into zones according to the guidelines outlined in Section~\ref{sect:segmentation-overview}, using the logical operator tool in 3D Slicer. The initial segmentations were reviewed and refined by an experienced medical imaging scientist (WS) and three surgery residents (JRK, WU, and MF), and were subsequently reviewed and edited by the expert vascular surgeon (MC) to establish the ground truth annotations. On average, the annotation of a single CTA image volume took approximately 10–15 hours.



\subsection{Evaluation Metrics and Ranking Criterion}
We evaluated model performance using two metrics: the Dice Similarity Coefficient (DSC) and the Normalized Surface Distance (NSD), also known as normalized surface Dice~\citep{seidlitz2022robust}. The DSC quantifies the overlap between the predicted segmentation mask (\(\text{P}\)) and the ground truth mask (\(\text{G}\)), ranging from 0 (no overlap) to 1 (perfect overlap). It is calculated as:
\[
\text{DSC} = \frac{2 \cdot |\text{P} \cap \text{G}|}{|\text{P}| + |\text{G}|},
\]
where \(|\cdot|\) denotes the cardinality (i.e., the number of elements in the set). The DSC balances precision and recall, making it a widely used metric for segmentation tasks.

The NSD complements the DSC by evaluating boundary alignment between the predicted and ground truth segmentations. It measures the proportion of boundary points within a specified tolerance distance (\(\tau\)), yielding values between 0 (no overlap) and 1 (perfect overlap). The NSD is given by:
\[
\text{NSD} = \frac{1}{2} \left( \frac{\sum_{p \in \text{S}_{\text{pred}}} \mathbb{I}(\text{d}(p, \text{S}_{\text{gt}}) \leq \tau)}{|\text{S}_{\text{pred}}|} + \frac{\sum_{q \in \text{S}_{\text{gt}}} \mathbb{I}(\text{d}(q, \text{S}_{\text{pred}}) \leq \tau)}{|\text{S}_{\text{gt}}|} \right),
\]
where \(\text{S}_{\text{pred}}\) and \(\text{S}_{\text{gt}}\) are the predicted and ground truth surfaces, respectively. \(\text{d}(p, \text{S}_{\text{gt}})\) represents the shortest distance from a point \(p\) on the predicted surface to the ground truth surface, and \(\text{d}(q, \text{S}_{\text{pred}})\) is defined similarly. \(\mathbb{I}(\cdot)\) is an indicator function (1 if the condition is true, 0 otherwise), and \(|\text{S}_{\text{pred}}|\) and \(|\text{S}_{\text{gt}}|\) denote the total number of points on the predicted and ground truth surfaces, respectively. In our computation, we set \(\tau = 2\).

Together, DSC and NSD provide a comprehensive assessment of segmentation performance, evaluating overall overlap and boundary precision.
To rank participating teams, we adopted a "rank-then-aggregate" strategy, inspired by recent segmentation challenges~\citep{ma2022fast, bakas2018identifying}. The process was as follows:

\begin{enumerate}
    \item For each finalist team, we applied their model to 40 testing images. For each CTA image, we calculated the DSC and NSD for each of the 23 labels, then averaged them to obtain the mean DSC and NSD. This resulted in 40 DSC values and 40 NSD values per team.
    \item For each CTA image, we ranked all teams from 1 to \(N\) (\(N\) being the total number of teams), assigning two rank values per team: one based on DSC and the other based on NSD.
    \item For each team, the 40 DSC-based rankings and 40 NSD-based rankings were averaged, resulting in the average Dice ranking and the average NSD ranking.
    \item For each team, we averaged their DSC and NSD rankings to compute an overall ranking score.
    \item Finally, we ranked all teams based on their overall ranking scores.
\end{enumerate}

\section{Results}
\label{sect:results}
\subsection{Challenge Participants and Submissions}
\label{sect:challenge-participants}
\begin{figure}[!hbt]
    \centering
    \includegraphics[width=\linewidth]{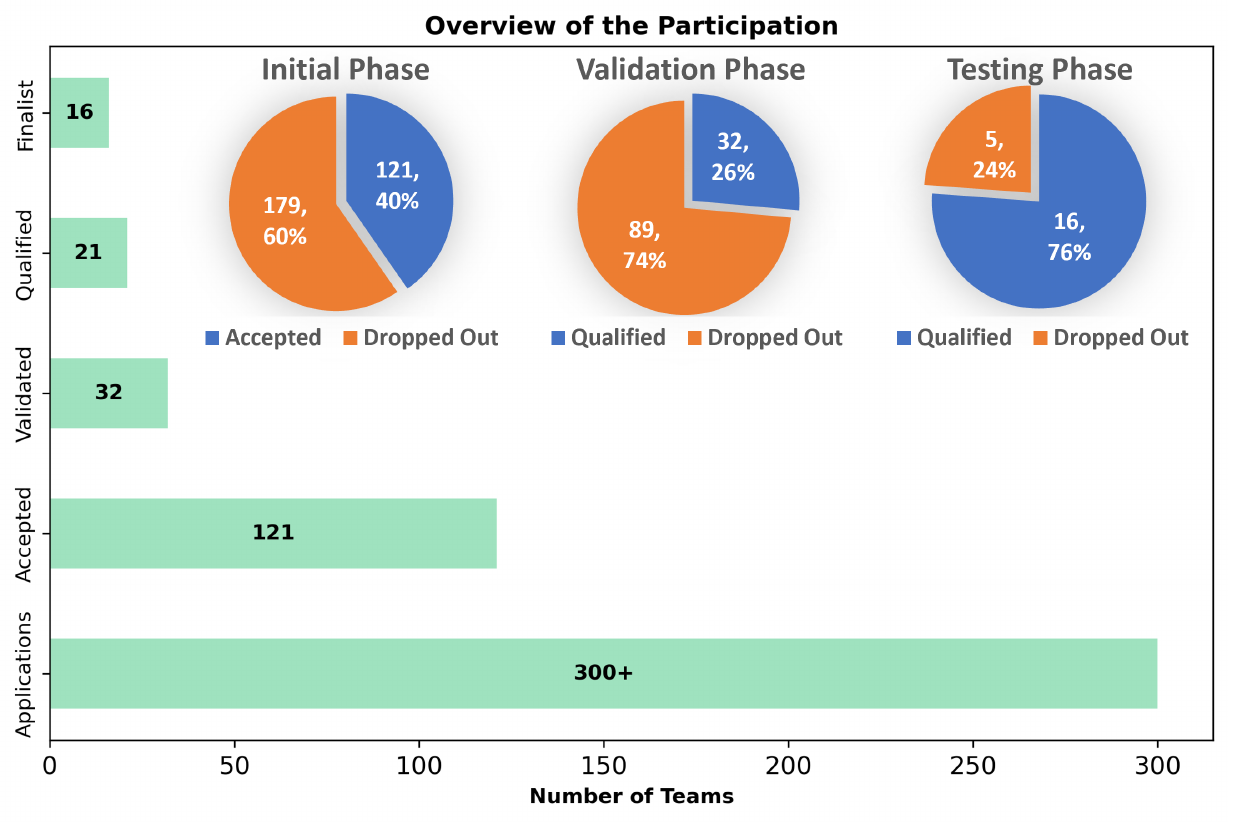}
    \caption{Overview of the participating teams at each phase.}
    \label{fig:participation-overview}
\end{figure}

The registration process for the AortaSeg24 challenge involved two main steps: joining the challenge on the Grand Challenge platform (\url{https://aortaseg24.grand-challenge.org}) and submitting a signed data use agreement form outlining the challenge's rules. Figure~\ref{fig:participation-overview} provides an overview of participant engagement across all phases of the challenge. Over 300 applications were received on the Grand Challenge platform, and following a manual review, 121 teams were approved. Participants were primarily declined if they failed to submit the signed data use agreement or did not have a valid affiliation. During the validation phase, 32 participating teams successfully submitted their Docker images, which were evaluated on ten hidden validation images. Teams could immediately view their resulting Dice scores, which needed to exceed the Dice score of the baseline 3D U-Net model (0.713 on the validation dataset) to qualify for the testing phase. Of the 32 teams that submitted Docker images, 21 met this threshold and advanced to the testing phase. In the testing phase, 16 valid submissions were received. The Grand Challenge platform evaluated each team's Docker image on 40 hidden testing images. All 16 finalist teams were invited to submit a short technical paper and co-author this challenge paper.

\begin{table*}[!h] 
    \caption{Key features of the top five algorithms. SGD: stochastic gradient descent. AdamW: Adam optimizer with decoupled weight decay.}
    \centering
    \renewcommand{\arraystretch}{0.2} 
    \resizebox{0.9\textwidth}{!}{
    \begin{tabular}{c | c | c | c | c | c | c | c}
        \toprule[0.8pt]
        Algorithm  & \makecell{Architecture} & \makecell{Data\\Augmentation} & \makecell{Loss\\Function}  & \makecell{\# of\\Epochs} & \makecell{Cross\\Validation} & \makecell{Post\\Processing} & \makecell{Optimizer} \\
        \midrule[0.8pt]
        \makecell{A1}  &   \makecell{Cascaded \\ 3D nnU-Net at \\ different resolutions}  & \makecell{Rotation\\ Scaling} & \makecell{Dice loss\\ + \\ cross-entropy loss}   & 800 & 5-fold & None & \makecell{SGD} \\
        \hline
        \makecell{A2}  &   \makecell{nnU-Net}  & \makecell{nnU-Net default\\ No flipping}& \makecell{Dice loss\\ +\\ cross-entropy loss \\ + \\ cbDice loss}   & \makecell{stage 1: 500\\stage 2: 500} & \makecell{stage 1: fold 0\\stage 2: 5-fold} & \makecell{Removal of small\\targets between\\ two stages} & \makecell{SGD} \\
        \hline
        \makecell{A3}  &   \makecell{nnU-Net\\ResEnc L}  & \makecell{nnU-Net DA5} & \makecell{Dice loss\\+\\cross-entropy loss}   &  \makecell{1000} & \makecell{5-fold} & \makecell{None} & \makecell{SGD} \\
        \hline
        \makecell{A4}  &   \makecell{nnU-Net}  & \makecell{Cropping\\Flipping\\Rotation\\Scaling\\Random noise} & \makecell{Dice loss\\+\\cross-entropy loss}  & \makecell{500} & \makecell{5-fold} & \makecell{Largest\\connected\\component} & \makecell{SGD} \\
        \hline
        \makecell{A5}  &   \makecell{SegResNet\\+\\nnU-Net \\ResEnc L}  & \makecell{Cropping\\Affine transformation\\ flipping} & \makecell{Dice loss\\+\\cross-entropy loss\\+\\skeleton recall loss}  & \makecell{2000} & \makecell{5-fold} & \makecell{None} & \makecell{AdamW} \\
        \bottomrule[0.8pt]
    \end{tabular}}
    \label{tab:top-five-algorithm-summary}                         	
\end{table*}

\subsection{Approaches of the Top Five Teams}
\label{sect:top-5-team-methods}
Table~\ref{tab:top-five-algorithm-summary} highlights the key features of the top five teams in the AortaSeg24 challenge. The algorithms are referred to by their final rankings, with the highest-ranked algorithm labeled as A1, the second as A2, and so on. For conciseness, we provide a brief summary of the approaches used by the top five teams. Detailed descriptions of each team's methodology can be found in their short technical papers.

\subsubsection{A1: The Best-Performing Algorithm}
The team designed an efficient cascaded framework that first localized the region of interest (ROI) followed by fine-grained segmentation on the full-resolution ROI. This approach begins with coarse segmentation at a spacing of 1.9$\times$1.9$\times$1.9 mm³. The fine-segmentation stage uses the image at full resolution, with a spacing of 1.0$\times$1.0$\times$1.0 mm³. The method ensembles the inference results from five models: three trained based on nnUNet with a ResEncM architecture and two trained based on nnUNet with a ResEncL architecture. The ResEncL and ResEncM architectures have more parameters and larger patch sizes compared to the previous 3D fullres architecture. The nnUNet is trained for 800 epochs, incorporating data augmentation techniques such as random rotation and scaling, though mirroring operations were not used. The binary cross-entropy (CE) loss and Dice loss were used as supervised loss functions to train the model. The network was trained on a single NVIDIA GeForce RTX 3090 Ti GPU with 24 GB memory, using the SGD optimizer and a decaying learning rate initialized at 0.01. The source code is publicly available at \url{https://github.com/MaxwellEng/MICCAI_CHANLLENGE24_HJL}.

\subsubsection{A2: The Second-Best-Performing Algorithm}
The team developed a two-stage segmentation framework to address the complexity of aortic vessel segmentation. Initially, the framework segments the binary foreground region of the aorta, followed by cropping and fine-grained segmentation of individual aortic classes. To manage the intricate vascular structures, a hierarchical semantic learning strategy inspired by human cognition was employed, progressively incorporating anatomical constraints by decomposing complex structures. This hierarchical approach led to faster training convergence and improved segmentation accuracy. The integration of hierarchical semantic loss resulted in a 5–15\% increase in the Dice coefficient from the early stages of training. Furthermore, the centerline boundary Dice (cbDice) loss function was employed to ensure consistent segmentation across varying vessel diameters. The two-stage inference process first segments all foreground vessels using a low-resolution 3D model, followed by multi-class segmentation within a refined region of interest (ROI). This significantly reduces inference time, accelerates the overall process by up to five times, and enhances the practicality of the model for real-time clinical applications. The code is available at: \url{https://github.com/PengchengShi1220/AortaSeg24}.

\subsubsection{A3: The Third-Best-Performing Algorithm}
The team, with their algorithm Hauptschlagader, utilized the nnU-Net ResEncL framework, achieving a top Dice score of 77.54 in cross-validation. Replacing the default nnU-Net configuration with the ResEncL framework allowed the model to better capture complex vascular structures by leveraging a larger patch size and, consequently, a higher receptive field. Additional improvements included disabling left/right mirroring to address anatomical asymmetry and employing the DA5 augmentation strategy to enhance robustness on small datasets. These optimizations significantly outperformed the default nnU-Net. The code is available at \url{https://github.com/MIC-DKFZ/nnUNet/blob/master/documentation/competitions/AortaSeg24.md}.

\subsubsection{A4: The Fourth-Best-Performing Algorithm}
The team developed a coarse-to-fine model for segmenting the aorta and its branches. The Coarse model first localized the approximate region of interest, while the Fine model performed detailed segmentation within the detected region. Both models were built on the nnU-Net framework and shared the same architecture. However, they differed in their outputs: the Coarse model was designed to determine whether a region contained foreground, while the Fine model segmented the foreground into 23 distinct classes. To enhance inference efficiency, the Coarse stage used a dynamic resolution approach, resampling input images into uniformly sized patches. In contrast, the Fine stage maintained a fixed resolution and employed a sliding-window strategy to ensure high segmentation accuracy. This sliding-window strategy was applied exclusively to foreground areas in the Fine stage, minimizing the influence of background regions and improving both accuracy and inference efficiency. The code is available at \url{https://github.com/LSL000UD/AortaSeg24/tree/main}.

\subsubsection{A5: The Fifth-Best-Performing Algorithm}
The team developed a Residual U-Net approach with two frameworks, nnU-Net \citep{isensee2021nnu} and MONAI \citep{cardoso2022monai}, to effectively utilize partially labeled data. Using the MONAI framework, they employed the SegResNet model, combining Dice Cross-Entropy Loss with Skeleton Recall Loss \citep{kirchhoff2024skeleton}. Deep supervision was also applied to guide the skeleton masks. For training preparation, they created skeleton masks from the original segmentation masks following the methodology outlined in the Skeleton Recall Loss paper, optimizing GPU and CPU performance during loss computation. Images, segmentation masks, and skeletons were used as inputs during training. With the nnU-Net framework, they utilized a Large Residual Encoder nnU-Net (ResEncL) model, also combining Dice Cross-Entropy Loss with deep supervision. Both the SegResNet and ResEncL models shared the same input size, data augmentation techniques, Z-score normalization, and optimizer settings. Although they attempted to train ResEncL with Skeleton Recall Loss, the results were less effective than their custom SegResNet implementation with Skeleton Recall Loss. Their SegResNet achieved a Dice score comparable to that of ResEncL. The code is publicly available at \url{https://github.com/BongNT/AortaSeg}.

\subsection{Rankings of Segmentation Algorithms}
The performance of 16 segmentation algorithms (A1 to A16) was evaluated across 40 testing subjects using two key metrics: Dice Similarity Coefficient (DSC) and Normalized Surface Distance (NSD), which measure shape overlap and boundary alignment, respectively. For each algorithm, a DSC ranking score was computed by averaging its ranks across the 40 testing cases, and the corresponding rank based on its DSC ranking score is presented in the DSC Rank column of Table~\ref{tab:overal-ranking}. Similarly, the NSD ranking score of each algorithm was averaged across 40 cases, with the resulting ranks shown in the NSD Rank column. The final ranking of each algorithm was determined by averaging its DSC and NSD ranking scores, and these overall rankings are displayed in the Final Rank column of Table~\ref{tab:overal-ranking}
\begin{table}[!h]
\centering
\caption{Rankings of 16 algorithms based on their DSC and NSD across 40 testing cases.}
\renewcommand{\arraystretch}{1.3} 
\scalebox{0.85}{
\begin{tabular}{c| c | c | c  |c |c }
\toprule[0.8pt]
Algorithms              & Avg DSC          & Avg NSD     &   \makecell{DSC\\Rank} &   \makecell{NSD\\Rank} &  \makecell{Final\\Rank} \\
\midrule[0.8pt]
A1  &   0.782 ± 0.025	&   0.817 ± 0.027   &   1   &   1	&   1      \\
A2  &   0.779 ± 0.028   &   0.812 ± 0.033	&   2   &   2	&   2      \\
A3	&   0.779 ± 0.025 	&   0.809 ± 0.027	&   3	&   3	&   3      \\
A4	&   0.772 ± 0.033	&   0.805 ± 0.037	&   4	&   4	&   4      \\
A5	&   0.773 ± 0.028	&   0.805 ± 0.038 	&   5	&   5	&   5      \\
A6	&   0.767 ± 0.032   &   0.797 ± 0.037	&   6	&   6	&   6      \\
A7	&   0.767 ± 0.029	&   0.795 ± 0.039 	&   7	&   8	&   7      \\
A8	&   0.766 ± 0.025 	&   0.798 ± 0.030    &   8	&   7	&   8      \\
A9	&   0.760 ± 0.033 	&   0.791 ± 0.042 	&   10	&   9	&   9      \\
A10	&   0.752 ± 0.052 	&   0.782 ± 0.055	&   9	&   10	&   10      \\
A11	&   0.755 ± 0.038	&   0.788 ± 0.042   &   11	&   11	&   11      \\
A12	&   0.752 ± 0.026 	&   0.785 ± 0.033	&   13	&   12	&   12      \\
A13	&   0.737 ± 0.080 	&   0.765 ± 0.087 	&   14	&   13	&   13      \\
A14	&   0.743 ± 0.050	&   0.770 ± 0.057 	&   12	&   14	&   14      \\
A15	&   0.729 ± 0.043 	&   0.751 ± 0.047	&   15	&   15	&   15      \\
A16	&   0.675 ± 0.072 	&   0.693 ± 0.079	&   16	&   16	&   16      \\
\bottomrule[0.8pt]
\end{tabular}}
\label{tab:overal-ranking}                           	
\end{table}

The results in Table~\ref{tab:overal-ranking} show that Algorithm A1 consistently outperformed all others, achieving the highest average DSC (0.782) and NSD (0.817), demonstrating its superior ability to capture both shape and boundary details accurately. A2 and A3 closely followed A1, with only marginal differences in average DSC (0.003 lower). The top five algorithms exhibited consistent rankings across both metrics, indicating a high degree of alignment between their shape and boundary-capturing capabilities. Among these, a closer examination of A4 and A5 reveals nuances in their evaluation. The average Dice values for A4 and A5 are 0.772 ± 0.033 and 0.773 ± 0.028, respectively, indicating that A5 slightly outperformed A4 in terms of average Dice. However, A4 achieves a better average ranking score based on Dice (5.75 vs. 5.8 for A5). This discrepancy highlights the difference between evaluating performance using metric averages and a rank-and-aggregate method.

In contrast to the top-performing algorithms, A15 and A16 ranked the lowest across all metrics, indicating difficulties in capturing shape overlap and aligning boundaries. Mid-tier algorithms, such as A6 to A8, demonstrated moderate performance. A comparison of A13 and A14 highlights nuances in evaluation: while A14 had slightly higher average DSC (0.743 vs. 0.737) and NSD (0.770 vs. 0.765) scores, A13 performed better in most cases, as shown by its superior ranking score (10.75 vs. 10.98). This underscores the importance of evaluating case-by-case performance rather than relying solely on averages. Overall, top-performing algorithms like A1, A2, and A3 are well-suited for high-precision tasks, while mid- and lower-ranked algorithms require further optimization for clinical or real-world applications.


\subsection{Analysis of the Distributions of Dice and NSD Scores}
\label{sect:dsc-and-nsd-analysis}

\begin{figure}[!hbt] 
\centering 
\includegraphics[width=0.48\textwidth]{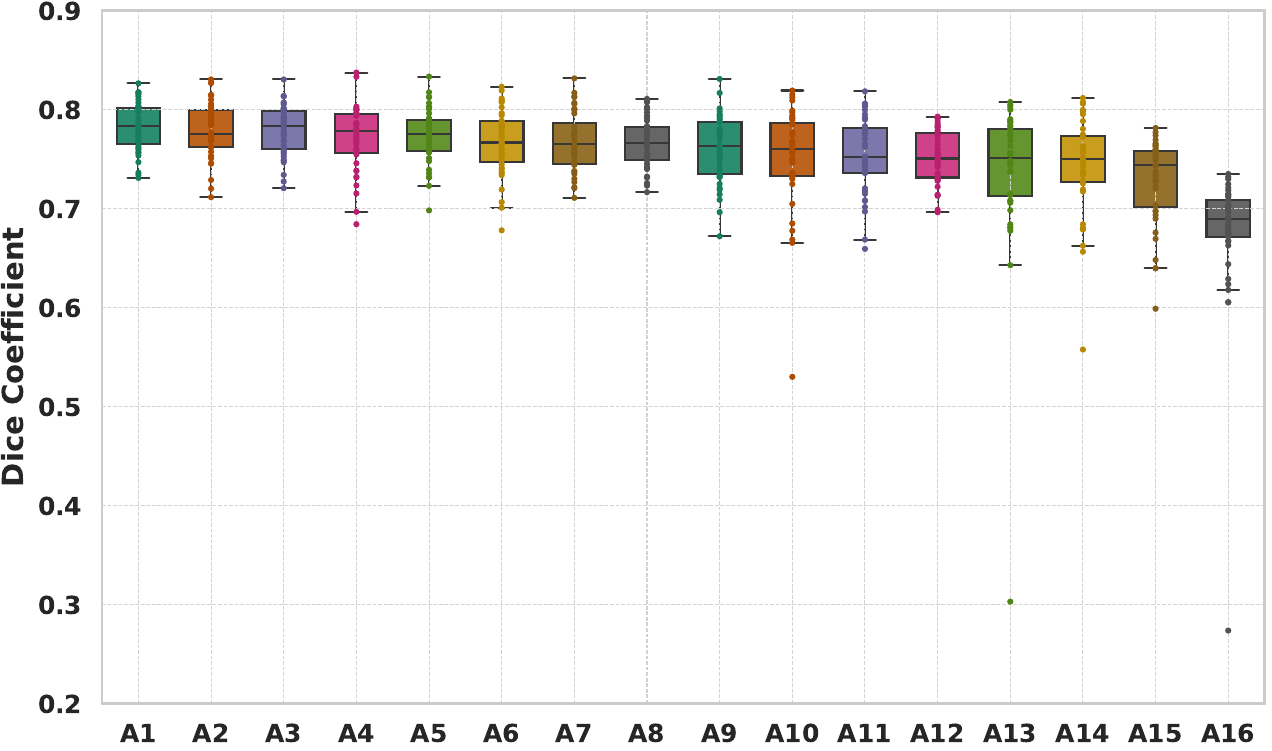} 
\caption{Box plot showing the distribution of Dice scores for 16 segmentation algorithms. The x-axis represents the team names, and the y-axis shows the Dice scores. Each dot corresponds to the Dice score for an individual subject segmented by an algorithm.
} 
\label{fig:box_and_dot_plot_for_dice} 
\end{figure}

The Dot-and-Box plot in Fig.~\ref{fig:box_and_dot_plot_for_dice} illustrates the Dice score distributions for each segmentation algorithm across the 40 testing cases. The top five algorithms (A1 to A5) achieved higher median Dice scores than all other algorithms, demonstrating their superior segmentation performance.
Among these, A1 had the highest median Dice score and a relatively narrow interquartile range (IQR), reflecting its robust and consistent performance. A2, despite having the second-best 25th and 75th percentiles, exhibited the lowest median Dice score among the top five. A4 outperformed A2 in terms of median Dice score but had two outliers with Dice scores below 0.7. The mid-tier algorithms (A6 to A10) also demonstrated promising performance, though A9 and A10 showed much wider IQRs and more outliers, indicating higher variability. The lower-performing algorithms (A11 to A16) exhibited lower medians and significantly more outliers, underscoring challenges in maintaining consistency.

\begin{figure}[h]
    \centering
    \includegraphics[width=0.48\textwidth]{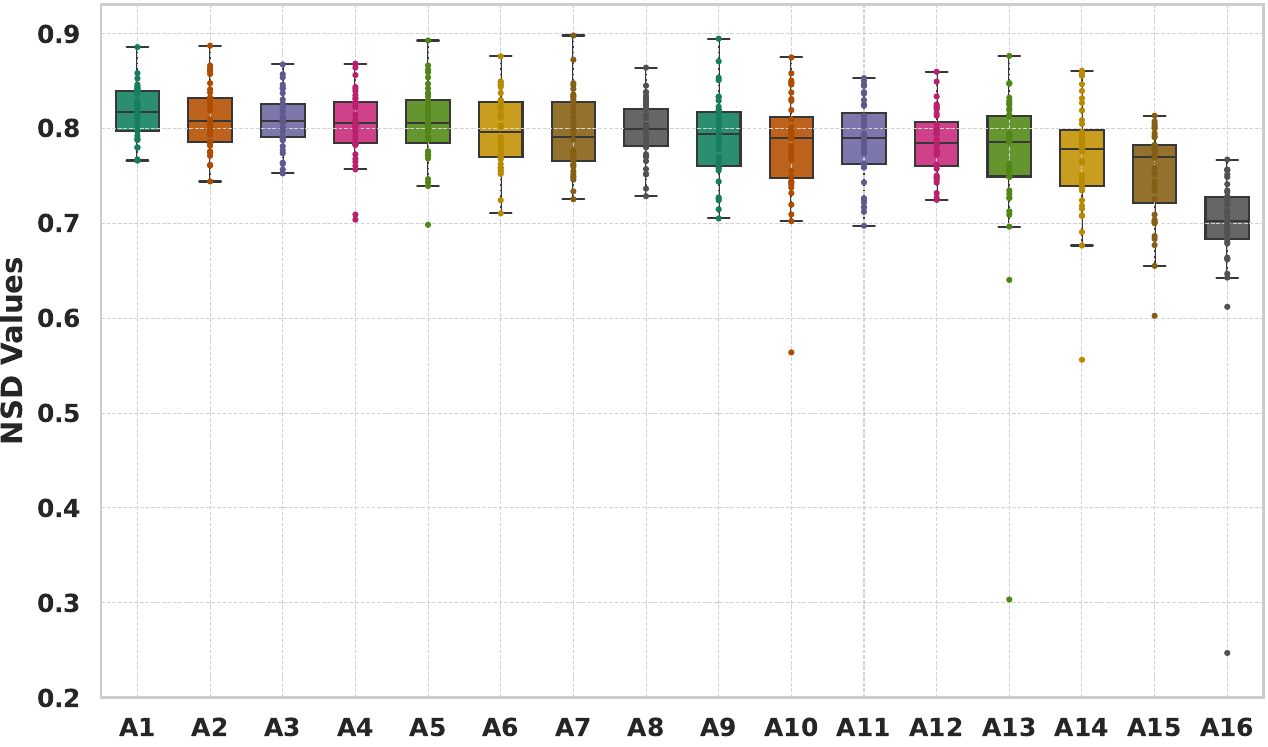}
    \caption{Box plot showing the distribution of NSD scores for 16 segmentation algorithms. The x-axis represents the team names, and the y-axis shows the NSD scores. Each dot corresponds to the NSD score for an individual subject segmented by an algorithm.}
    \label{fig:box_and_dot_plot_for_nsd}
\end{figure}

The Dot-and-Box in Fig.~\ref{fig:box_and_dot_plot_for_nsd} illustrates the NSD score distributions for each segmentation algorithm. The top five algorithms achieved higher median NSD scores than all other algorithms, highlighting their superior boundary alignment capabilities. 
A1 had the highest median NSD score and the second narrowest IQR among the top five algorithms. A2 and A3 showed comparable median NSD scores, with A2 having a higher 75th percentile and A3 exhibiting a slightly narrower IQR. The mid-tier algorithms (A6 to A10) also demonstrated strong performance but exhibited wider IQRs (except for A8) and more outliers compared to the top five. The lower-performing algorithms (A11 to A16) displayed noticeably lower median NSD scores and a higher number of outliers. These results in Fig.\ref{fig:box_and_dot_plot_for_dice} and Fig.\ref{fig:box_and_dot_plot_for_nsd} underscore the robust performance of the top-performing algorithms while highlighting the variability and challenges faced by mid-tier and lower-ranked algorithms.

\subsection{Qualitative Analysis of Segmentation Results}
\begin{figure}[!hbt]
    \centering
    \includegraphics[width=\linewidth]{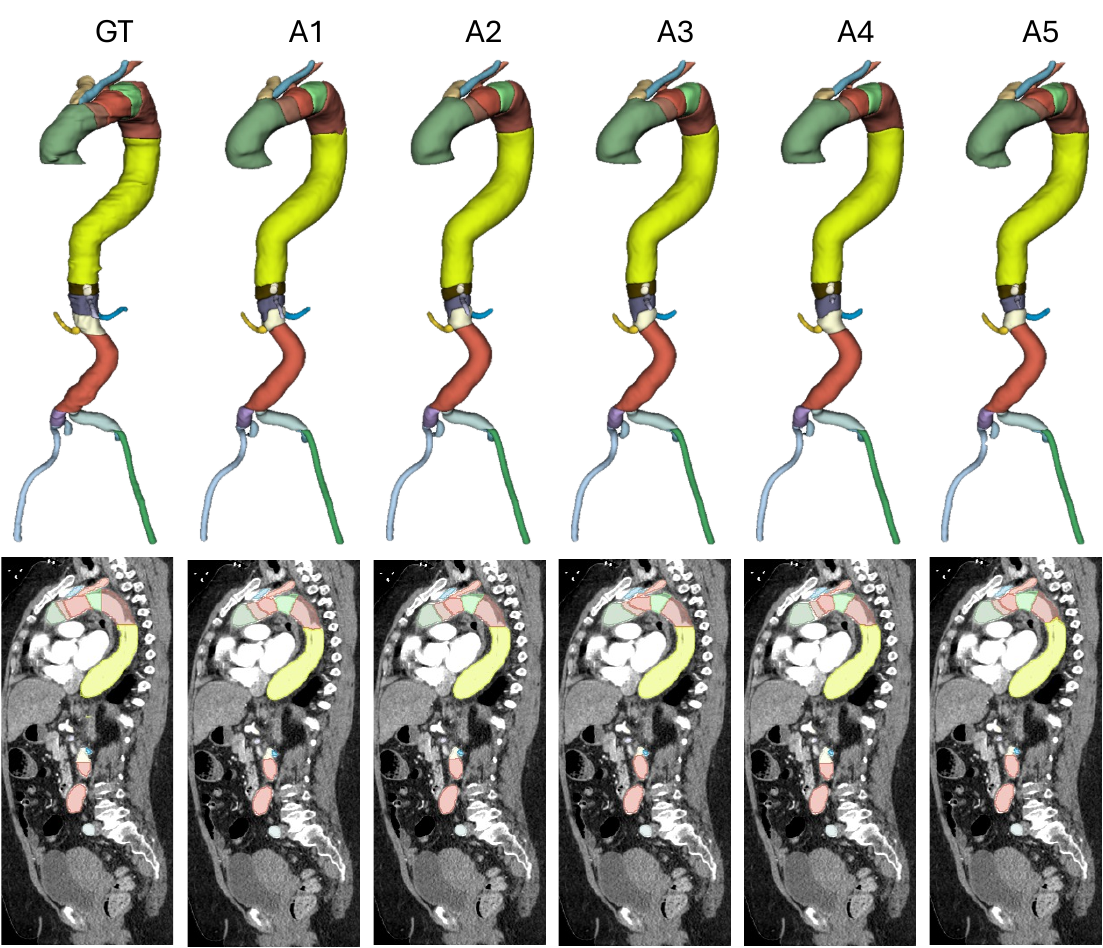}
    \caption{Qualitative results for a challenging case, showing 3D outputs and sagittal slices. The ground truth (first column) is compared with results from the top five algorithms (A1–A5).}
    \label{fig:visual_results_comparison}
\end{figure}

Figure~\ref{fig:visual_results_comparison} shows segmentation results for a representative case. The top row presents 3D renderings of the ground truth and segmentations from the top five teams, while the bottom row shows a sagittal CT slice overlaid with their segmentations. The top five algorithms achieved high accuracy, with minimal over-segmentation or under-segmentation and only slight differences in their results. All methods used the same architecture, nnU-Net~\citep{isensee2021nnu}. This case is particularly challenging due to the left common carotid artery’s close proximity to zone 0 and the innominate artery. The top two algorithms (A1 and A2) partially resolved this, while the remaining three incorrectly identified the left common carotid artery as originating from the innominate artery instead of zone 1. Overall, the top five algorithms demonstrated exceptional accuracy in segmenting the aorta, its branches, and its zones.

\section{Discussion}

\subsection{Significance of the AortaSeg24 Challenge}

The AortaSeg24 challenge represents a major advancement in aortic disease management by introducing multi-class segmentation of the aorta, enabling precise measurements of diameters and volumes across different zones. These anatomical metrics are essential for guiding clinicians in treatment selection and planning endovascular procedures like stent graft placement. According to ACC/AHA guidelines, the risk of aortic dissection rises sharply at diameters $\geq$4.5 cm~\citep{writing20222022}, and earlier intervention can improve patient outcomes. Automating aortic segmentation reduces variability and the labor-intensive process of manual annotations, enhancing care quality. Furthermore, by streamlining aortic measurements, AortaSeg24 enables large-scale longitudinal data analysis, improving predictions of disease progression and rupture risk while supporting better clinical decision-making.

Another major contribution of the AortaSeg24 challenge is the release of 100 publicly available CTA images with expert annotations. As the largest publicly available dataset for multi-class segmentation of aortic structures, it addresses a critical barrier in medical image analysis: the scarcity of comprehensive, high-quality annotated datasets. The inclusion of expert annotations enhances the dataset's clinical relevance, making it a robust foundation for developing real-world aortic segmentation algorithms. With 23 distinct aortic branches and zones, the dataset also serves as a valuable benchmark for evaluating general-purpose 3D image segmentation methods.

\subsection{Key Strategies of Top-Performing Teams}

The top-performing teams in the AortaSeg24 challenge employed innovative strategies to achieve high segmentation accuracy, surpassing the current state-of-the-art CIS-UNet model~\citep{imran2024cis,krebs2024volumetric}. These strategies commonly included the use on the nnU-Net architectures, extensive data augmentation (e.g., rotation, scaling, and cropping), Dice-based and task-specific loss functions, and cross-validation. Most teams relied on stochastic gradient descent (SGD) for optimization, with one exception (A5) utilizing the AdamW optimizer.

The best-performing algorithm (A1) implemented a cascaded framework combining coarse and fine-grained segmentation, leveraging ResEncL and ResEncM variants of nnU-Net with Dice and binary cross-entropy (BCE) loss. Similarly, A2 designed a two-stage nnU-Net framework for binary-to-multi-class segmentation, incorporating the centerline boundary Dice (cbDice)~\citep{shi2024centerline} loss for better boundary accuracy. A3 enhanced the nnU-Net ResEncL architecture with larger patch sizes and extensive data augmentation to effectively capture complex structures. A4 applied a coarse-to-fine nnU-Net approach, refining segmentations with a sliding-window strategy for high precision. A5 uniquely integrated SegResNet and nnU-Net, combining Skeleton Recall Loss~\citep{kirchhoff2024skeleton} with Dice and BCE loss to handle partially labeled data.

\subsection{Analysis of Ranking Stability}
\begin{figure}[!h]
        \centering
        \begin{subfigure}[t]{0.235\textwidth}
            \includegraphics[width=0.98\textwidth]{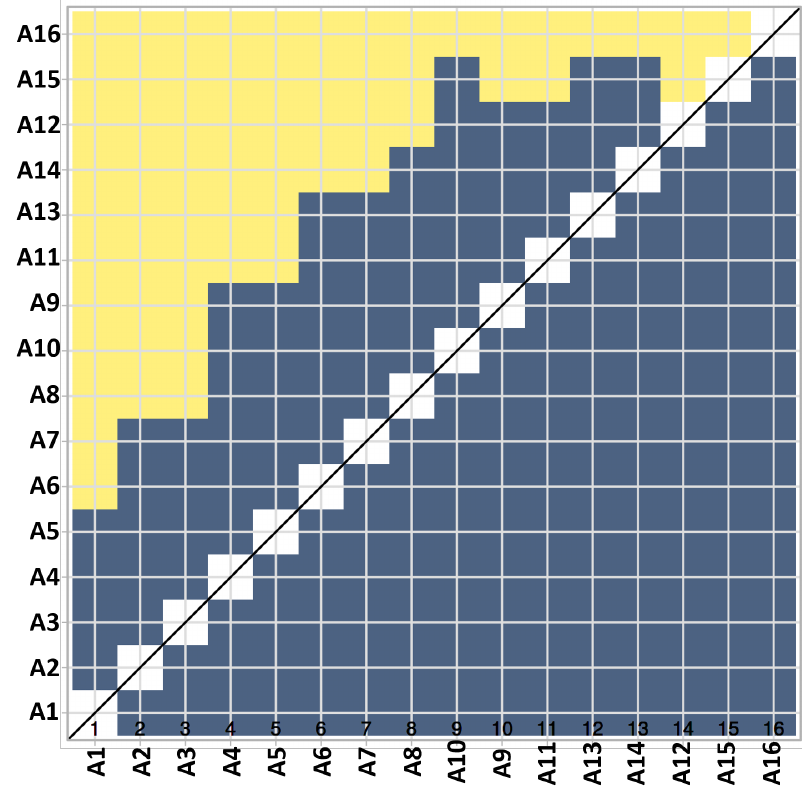}
            \caption{DSC Significance Map}
            \label{fig:dsc_significance_map}
        \end{subfigure}
        \begin{subfigure}[t]{0.235\textwidth}
            \includegraphics[width=0.98\textwidth]{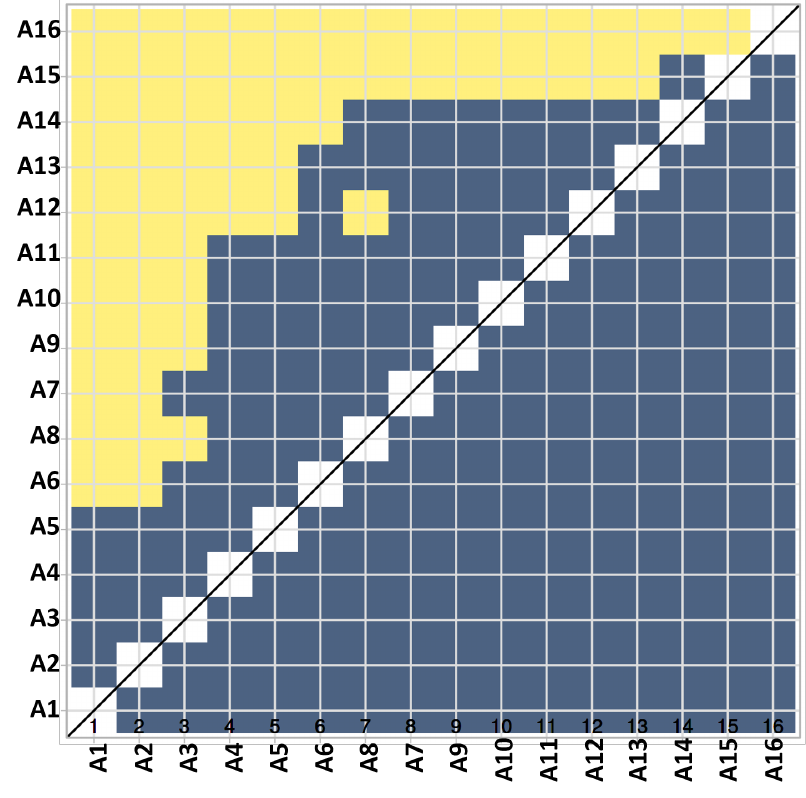}
            \caption{NSD Significance Map}
            \label{fig:nsd_significance_map}
        \end{subfigure}
\end{figure}

To evaluate the stability of rankings, we used significance maps as introduced by~\citep{wiesenfarth2021methods}, which graphically represent ranking stability through statistical significance. These maps show pairwise one-sided Wilcoxon signed-rank test results at a 5\% significance level, adjusted for multiple testing using Holm’s method. Figure~\ref{fig:dsc_significance_map} illustrates the significance map for Dice scores, where yellow shading indicates that the algorithm on the x-axis significantly outperformed the one on the y-axis, and blue shading indicates no significant difference. The results show that the top five algorithms (A1 to A5) consistently outperformed the others. For instance, the highest-ranked algorithm, A1, demonstrated significant superiority over eleven other algorithms (A6 to A16), while A2 and A3 significantly outperformed nine algorithms (A8 to A16). No significant differences were observed among the top five algorithms or within the mid-tier group (A6 to A10). A similar pattern is found in the NSD significance map, as shown in Figure~\ref{fig:nsd_significance_map}. The consistent trends across both DSC and NSD significance maps confirm the robustness of the evaluation metrics and reinforce the reliability of the ranking methodology.

\begin{figure}[!h]
    \centering
    \includegraphics[width=0.85\linewidth]{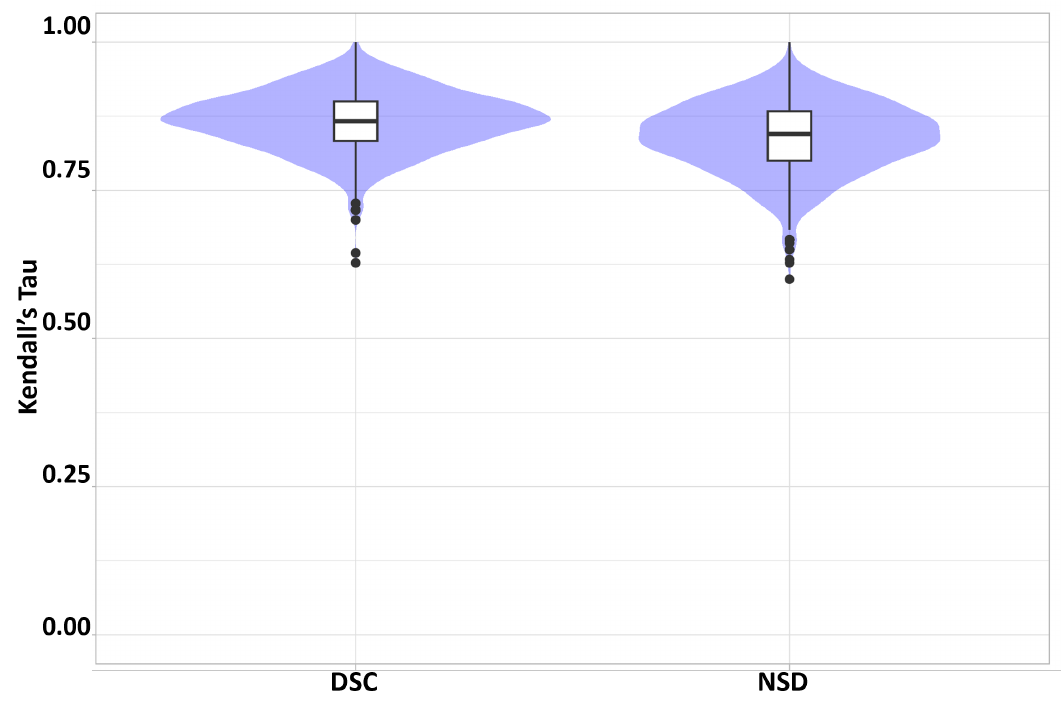}
    \caption{Violin plots for visualizing ranking stability based on bootstrapping in terms of DSC and NSD.}
    \label{fig:violinPlot}
\end{figure}

As another approach to evaluate ranking stability, we used bootstrap sampling to generate resampled datasets by drawing samples with replacement from the original data. This approach provided a distributional characterization of the estimates of ranking variability over 1,000 iterations. Kendall’s $\tau$, a correlation metric, was calculated for each pair of rankings derived from the bootstrap samples, with results visualized using violin plots in Fig.~\ref{fig:violinPlot}. Violin plots, which combine boxplot and density plot features, effectively summarize the distribution of Kendall’s $\tau$ values and offer a clear view of ranking correlations. For both DSC and NSD metrics, the violin plots demonstrate stable rankings across bootstrap samples, with relatively small variability. A narrow density and high Kendall’s $\tau$ values indicate strong ranking stability, affirming the reliability of DSC and NSD metrics. While significance maps highlight pairwise algorithmic superiority, violin plots complement them by providing insights into broader ranking reliability, ensuring that observed differences are not artifacts of sampling variability. Together, these methods create a robust evaluation framework that captures both the statistical significance of pairwise comparisons and the overall stability of algorithm rankings, offering a nuanced and comprehensive assessment of performance.

\subsection{Limitations and Future Directions}

The AortaSeg24 challenge has certain limitations that highlight areas for further improvement. First, the dataset is limited to uncomplicated type B aortic dissections (uTBAD), which restricts the applicability of the algorithms to other aortic conditions, such as type A dissections or aneurysms. Second, the reliance on a single imaging modality, computed tomography angiography (CTA), and consistent imaging protocols may reduce the generalizability of the algorithms to different clinical settings. Third, although expert annotations were used, inter-observer variability and difficulties in segmenting smaller aortic branches may have introduced inconsistencies, potentially affecting the accuracy of the models in capturing detailed anatomical structures. 

Future work will focus on addressing these limitations while incorporating clinically relevant tasks. Expanding the dataset to include cases with diverse aortic pathologies, imaging modalities such as 4D flow MRI, and data from multiple institutions will improve the generalizability of the algorithms. Additionally, the scope of the challenge can be broadened to include tasks critical to clinical decision-making, such as measuring the cross-sectional diameter of the aorta in each aortic zone, a key metric for assessing disease progression and guiding intervention planning. Finally, developing computationally efficient models will be prioritized to ensure accessibility across a range of clinical environments, including those with limited resources.

\section{Conclusion}
The AortaSeg24 Challenge has set a new benchmark for multi-class segmentation of the aorta by introducing the first large-scale dataset annotated for 23 clinically relevant branches and zones. This initiative enables precise and automated segmentation, supporting improved diagnosis, treatment planning, and monitoring for aortic diseases. The top-performing algorithms demonstrated exceptional accuracy and robustness, with potential for direct clinical application. By fostering innovation and collaboration, the AortaSeg24 Challenge establishes a strong foundation for advancing AI-driven solutions in vascular imaging, enhancing both research and clinical outcomes.

\section*{Conflict of interest statement}
The authors have no conflict of interest to declare.

\section*{Declaration of generative AI and AI-assisted technologies in the writing process}
During the preparation of this work the authors used the ChatGPT-4.0 tool in order to improve the readability and language of our paper. After using this tool/service, the authors reviewed and edited the content as needed and take full responsibility for the content of the publication.

\section*{Acknowledgments}
This work was supported by the Department of Medicine and the Intelligent Clinical Care Center at the University of Florida College of Medicine.

\bibliographystyle{model2-names.bst}\biboptions{authoryear}
\bibliography{refs}

\end{document}